\documentclass[twoside]{ilcws07}
\usepackage[latin1]{inputenc}
\usepackage[dvips]{graphicx,epsfig,color}
\usepackage{wrapfig,rotating}
\usepackage{amssymb,amsmath,array}

\begin{document}
\title{\vspace*{-1.8cm}
\begin{flushright}
{\normalsize\bf LAL 07-150}\\
\vspace*{-0.2cm}
{\normalsize November 2007}
\end{flushright}
\vspace*{2cm}
New Analysis of SUSY Dark Matter Scenarios at ILC~\footnote{Contribution to 
LCWS07. The original title of the contribution is ``SM Background 
Contributions Revisited for SUSY Dark Matter Stau Analyses''}
\vspace*{1cm}} 

\author{\bf Zhiqing Zhang
\vspace{.3cm}\\
LAL, Univ\ Paris-Sud, CNRS/IN2P3, Orsay, France
\vspace{1cm}\\
}

\maketitle

\begin{abstract}
Applying realistic veto efficiencies for the low angle electromagnetic 
calorimeter located in the very forward direction of the future 
international linear collider, we revisited the Standard Model background 
contributions studied previously in stau analyses with supersymmetrical 
dark matter scenarios.
\end{abstract}
\vspace*{1cm}

In supersymmetry (SUSY) models with $R$-parity conservation, 
the lightest SUSY particle neutralino, $\tilde{\chi}^0_1$, is often considered 
as the best candidate to satisfy the cosmological constraints on 
cold Dark Matter (DM) of the universe. 

In two previous studies~\cite{bbrz04,um04}, one of the most challenging 
scenarios analyzed concerns the benchmark point $D^\prime$~\cite{ellis03}
in the so-called co-annihilation region. 
In the mSUGRA model, the mass spectrum depends on two parameters $m_0$ and 
$M_{1/2}$, the common masses of scalars and gauginos superpartners at 
the unification scale. The parameter $\mu$, defining the higgsino mass, 
is derived, in absolute value, by imposing the electroweak symmetry breaking 
condition in terms of these two parameters and of $\tan\beta$, 
the ratio of the vacuum expectations which appear in the two Higgs doublets of
SUSY.
In scenario $D^\prime$, these parameters take the value $m_0=101\,{\rm GeV}$,
$M_{1/2}=525\,{\rm GeV}$, $\tan\beta=10$ and sign$(\mu)<0$. 
The resulting $\tilde{\chi}^0_1$
has a mass value of $212\,{\rm GeV}$ and the next lightest 
SUSY particle stau, $\tilde{\tau}$, has a mass value of $217\,{\rm GeV}$.
The mass difference is only $5\,{\rm GeV}$. 
When the mass difference is small, the co-annihilation process 
$\tilde{\chi}^0_1 \tilde{\tau}\rightarrow \tau\gamma$ becomes 
the dominant process for regulating the relic
DM density of the universe. It is therefore crucial to
measure precisely the mass values of $\tilde{\chi}^0_1$ and $\tilde{\tau}$.

The $\tilde{\chi}^0_1$ mass
can be measured~\cite{um04} using the end-point method with a precision 
down to $170\,{\rm MeV}$ ($80\,{\rm MeV}$) relying on 
$e^+e^-\rightarrow \tilde{\mu}^+\tilde{\mu}^-\rightarrow \mu^+\tilde{\chi}^0_1\mu^-\tilde{\chi}^0_1$ 
($\tilde{e}^+\tilde{e}^-\rightarrow e^+\tilde{\chi}^0_1e^-\tilde{\chi}^0_1$) 
for the modified SPS $1a$
scenario with a mass value of $\tilde{\mu}$ or $\tilde{e}$ of
$143\,{\rm GeV}$ and $\tilde{\chi}^0_1$ of $135\,{\rm GeV}$ under the following
experimental conditions: a center-of-mass energy (Ecm) of $400\,{\rm GeV}$,
an integrated luminosity (${\cal L}$) of $200\,{\rm fb}^{-1}$ and 
a polarized electron (positron) beam at $0.8$ ($0.6$). 

The stau analyses are more challenging not only because the final state 
particle of the tau decay is very soft with missing energy
due to undetected neutrino(s) in addition to $\tilde{\chi}^0_1$
but also because the Standard Model (SM)
background processes have rates which are many orders of magnitude larger
than that of the signal. The cross section values of the signal and
the dominant SM background processes are given in Table~\ref{tab:xsections}.
The signal row with Ecm$=442\,{\rm GeV}$ corresponds to the optimal 
center-of-mass energy method (referred to hereafter as {\em method one} 
using the cross section measurement or event counting near threshold)  
proposed  in~\cite{bbrz04} whereas the other signal rows correspond to cases
studied in another method ({\em method two} relying on the measured energy 
spectra of the tau decay final state, the first and other rows
are respectively studied in~\cite{um04} and~\cite{zz06}). 
\begin{table}
\centerline{\begin{tabular}{|c|c|c|}
\hline
Ecm (${\rm GeV}$) & Beam polarization ($P_{e^-}/P_{e^+}$) & $\sigma$ (fb) \\\hline
\multicolumn{3}{|c|}{Signal} \\ \hline
$600$ & $0.8/0.6$ & $50$ \\
$600$ &  unpolarized & $20$ \\
$500$ & $0.8/0.6$ & $25$ \\
$500$ & unpolarized & $10$ \\
$442$ & unpolarized & $0.456$ \\\hline
\multicolumn{3}{|c|}{Dominant SM backgrounds} \\\hline
$500$ & unpolarized & $4.3\cdot 10^5 (e^+e^-\rightarrow \tau^+\tau^-e^+e^-)$ \\
      &        & $8.2\cdot 10^5 (e^+e^-\rightarrow c\bar{c}e^+e^-)$ \\\hline
\end{tabular}}
\caption{Cross section values of the signal ($e^+e^-\rightarrow \tilde{\tau}^+\tilde{\tau}^-$) and the dominant SM background processes for different 
Ecm and beam polarizations.}
\label{tab:xsections}
\end{table}

The suppression of the dominant SM 
background processes
$e^+e^-\rightarrow \tau^+\tau^-e^+e^-$, $c\bar{c}e^+e^-$ depends critically
on whether the spectator $e^+$ and/or $e^-$ can be found in the low angle 
calorimeter (BeamCal) located at $370\,{\rm cm}$
from the interaction point in the very forward 
direction around the beam pipe. In the previous studies~\cite{bbrz04,um04},
either an ideal veto or an old realistic veto~\cite{as04} was assumed.

In this analysis, we revisit the SM background suppression using realistic
veto efficiencies obtained in a recent study~\cite{pdl06}. 
In this study, the BeamCal design is different for the small 
($0$ or $2\,{\rm mrad}$) or large ($20\,{\rm mrad}$) crossing angle
beam configuration. In the small crossing angle case, the BeamCal
has an inner (outer) radius of $1.5\,{\rm cm}$ ($16.5\,{\rm cm}$).
In order to identify an energetic spectator $e^+$ or $e^-$ out of several
TeV energy deposit from huge
number of low energy $e^+e^-$ pairs stemming from beamstrahlung photon 
conversions, 
the BeamCal is designed to have fine granularity and large
longitudinal segmentation. The resulting veto efficiency is about $100\%$
for high energy electrons close to the beam energy ($250\,{\rm GeV}$),
decreases down to $20\%$ for a $75\,{\rm GeV}$ electron near the inner
side of the calorimeter and is assumed to be fully inefficient for electrons 
below $75\,{\rm GeV}$.\newline
\vspace*{-0,8cm}
\begin{wrapfigure}{r}{0.5\columnwidth}
\vspace{-8mm}
\centerline{\includegraphics[width=0.475\columnwidth]{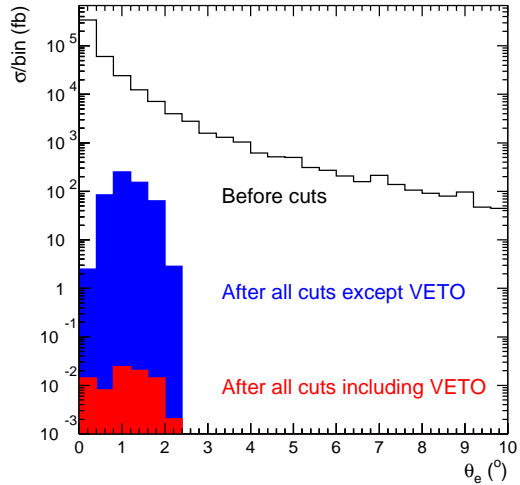}}
\caption{Angular distribution of the spectator electrons from 
$e^+e^-\rightarrow \tau^+\tau^- e^+e^-$ expressed in fb/bin. 
The blue shaded distribution corresponds to the distribution
obtained after all the selections described in~\cite{bbrz04} with 
the exception of the forward veto and the red shaded distribution
corresponds to the distribution when the veto is further included.}
\label{fig:veto}
\end{wrapfigure}

Taking the background process $e^+e^-\rightarrow \tau^+\tau^-e^+e^-$ as an
example, after applying all analysis cuts of {\em method one} defined 
in~\cite{bbrz04}, the remaining background amounts to $0.08\,{\rm fb}$ 
($561\,{\rm fb}$) when the forward veto is included (excluded).
This is illustrated in Fig.~\ref{fig:veto}.
This should be compared with the final signal cross section of $0.456\times
5.7\%=0.026\,{\rm fb}$ taking into account of the efficiency of the analysis.
The corresponding numbers for {\em method two} are $0.26\,{\rm fb} 
($168\,{\rm fb} without the veto) for the two-photon
$\tau^+\tau^-$ background 
and $10\times 6.4\%=0.64\,{\rm fb}$ for the signal at 
Ecm$=500\,{\rm GeV}$ and also with unpolarized beams.
The signal over background ratios for {\em method one} and {\em method two} 
are respectively $0.3$ and $2.5$. 
Therefore for {\em method one} where one is aiming
for a background free selection, the current veto and analysis selections
are not good enough and need further improvement.

\begin{figure}[h]
\vspace{-6mm}
\centerline{\includegraphics[width=\columnwidth]{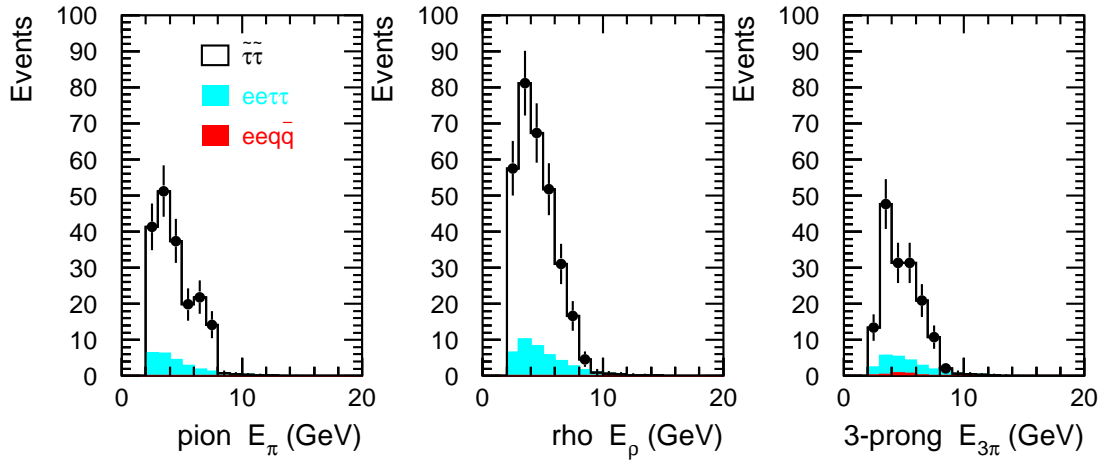}}
\vspace{-4mm}
\caption{The energy spectra of the hadronic final state in 
$\tau\rightarrow \pi\nu_\tau$, $\tau\rightarrow \rho\nu_\tau$ and 
$\tau\rightarrow 3\pi\nu_\tau$ decays from the signal reaction
$e^+e^-\rightarrow \tilde{\tau}^+\tilde{\tau}^-\rightarrow\tau^+\tilde{\chi}^0_1\tau^-\tilde{\chi}^0_1$ and two-photon production assuming
head-on collision and Ecm$=500\,{\rm GeV}$, ${\cal L}=300\,{\rm fb}^{-1}$
and $P_{e^-}=0.8$ and $P_{e^+}=0.6$.}
\label{fig:entau}
\end{figure}

For {\em method two}, although the absolute remaining background is
larger than that from {\em method one}, the background level is already 
acceptable, given the much bigger signal production cross section for
an Ecm well beyond the mass threshold.
In particular the signal over background ratio 
can substantially improve when the beams are polarized. 
This is shown in Fig.~\ref{fig:entau}.

Experimentally, the maximum $\tau$ energy ($E_{\rm max}$) can be determined 
from the upper end-point of the spectra, after having subtracted the small
SM background contribution, 
from a fit using for instance a polynomial function.
Since the maximum $\tau$ energy depends on Ecm, the mass values of 
$\tilde{\tau}$, $\tilde{\chi}^0_1$ and $\tau$, knowing $E_{\rm max}$, Ecm, 
$m_{\tilde{\chi}^0_1}$ and $m_\tau$ will thus allow one to derive the mass 
value of $\tilde{\tau}$. 
Assuming conservatively a precision of $100\,{\rm MeV}$
for the $\tilde{\chi}^0_1$ mass measurement, the $\tilde{\tau}$ mass
is expected to be measured in the range of $0.13-0.2\,{\rm GeV}$.
This in turn will result in an uncertainty of the DM density of 
$1.7-2.6\%$ based on the microMegas program~\cite{micromegas}.

The results for the benchmark scenario $D^\prime$ are summarized in 
Table~\ref{tab:result}. For the result of {\em method one}, we have assumed
that the background-free selection could be eventually achieved. 
The methods can also be applied to other co-annihilation scenarios. 
In general, the larger the mass difference between $\tilde{\tau}$ and 
$\tilde{\chi}^0_1$ is, the better the precision on the DM density will 
be~\cite{bbrz04,um04}.

\begin{table}[h]
\centerline{\begin{tabular}{|c|c|c|c|c|c|c|}
\hline
Ecm (${\rm GeV}$) & $P_{e^-}/P_{e^+}$ & ${\cal L}\,({\rm fb}^{-1})$ & $\sigma$ (fb) & Efficiency ($\%$) & $\delta m_{\tilde{\tau}}\,({\rm GeV})$ & $\delta \Omega h^2\,(\%)$ \\\hline
$600$ & $0.8/0.6$    & $300$ & $50$ & $7.6$ & $0.11-0.13$ & $1.4-1.7$ \\
$600$ &  unpolarized & $300$ & $20$ & $7.7$ & $0.14-0.17$ & $1.8-2.2$ \\
$500$ & $0.8/0.6$    & $300$ & $25$ & $6.4$ & $0.13-0.20$ & $1.7-2.6$ \\
$500$ & unpolarized  & $500$ & $10$ & $6.5$ & $0.15$      & $1.9$ \\
$442$ & unpolarized  & $500$ & $0.456$ & $5.7$ & $0.54$ & $6.9$ \\\hline
\end{tabular}}
\caption{Experimental conditions (Ecm, the beam polarizations and 
the integrated luminosity) and the corresponding results (the analysis 
efficiency, the stau mass uncertainty and the relative uncertainty on the
DM density determination).}
\label{tab:result}
\end{table}
In summary, we have revisited the SM background contributions to 
the challenging stau scenarios using the realistic veto efficiencies obtained
recently. If these scenarios are close to the one realized in nature, 
the uncertainty on the relic DM density obtained in linear collider 
can well match the precision to be expected from the Planck mission.

\section*{Acknowledgments}

The author wishes to thank P.~Bambade, M.~Berggren, F.~Richard for fruitful
collaboration, U.~Martyn for discussions and V.~Drugakov for proving the 
veto efficiency function.


\begin{footnotesize}



%

\end{footnotesize}


\end{document}